\documentclass[aps,prb,twocolumn,groupedaddress]{revtex4}
\usepackage{epsfig}
\begin{document}

\title{Structural dichroism in the antiferromagnetic insulating phase of V$_{2}$O$_{3}$. }
\author{C. Meneghini$^{1}$, S. Di Matteo$^{1,2}$,  C. Monesi$^{1}$, T. Neisius$^3$, L. Paolasini$^3$, S. Mobilio$^{1,2}$, C. R. Natoli$^{2}$, P.A. Metcalf$^4$, J.M. Honig$^4$ }
\affiliation {$^{1}${Dipartimento di Fisica "E. Amaldi", Universit\`a di Roma 3, via della Vasca Navale 84, I-00146 Roma, Italy}\\
$^{2}${Laboratori Nazionali di Frascati INFN, via E. Fermi 40 I-00044
Frascati, Roma, Italy}\\
$^{3}${European Synchrotron Radiation Facility, F-38043, Grenoble, France}\\
$^4$ Department of Chemistry, Purdue University, West Lafayette, IN47907, USA}
\date{\today}

\begin{abstract}

We performed near-edge x-ray absorption spectroscopy (XANES)  at V
K edge in the antiferromagnetic insulating (AFI) phase of a 2.8 $\%$
Cr-doped V$_2$O$_3$ single crystal. Linear dichroism of several percent is measured
in the hexagonal plane and found to be in good agreement
with ab-initio calculations based on multiple scattering theory. This experiment definitively proves the structural origin of the signal and therefore solves a controversy raised by previous interpretations of the same dichroism as non-reciprocal. It also calls for a
further investigation of the role of the
magnetoelectric annealing procedure in cooling to the AFI phase.

\end{abstract}

\maketitle

\section{Introduction}

In the last 35 years V$_2$O$_3$ has been the subject of wide
experimental and theoretical investigations,  after its
identification as the prototype of Mott-Hubbard systems.
\cite{rice} During the seventies an intensive set of
measurements\cite{mcwhan1,dernier,moon} was
performed in order to clarify its crystal and magnetic structure
as well as its transport properties.  Two metal-insulator
transitions were found: one from a paramagnetic metallic (PM) to
an antiferromagnetic insulating (AFI) phase, below $\simeq 150$ K,
associated with a corundum-to-monoclinic crystal distortion; the
other from the PM to a paramagnetic insulating (PI) phase at
higher temperatures (above $\simeq 500$ K). The PI phase can also be
obtained by means of a small ($1\div 2 \%$) Cr-doping: in
this case Cr ions act as a negative pressure, thus enlarging the
average  cation-cation distance and inducing a Mott-insulator
transition. As reviewed, eg, in Ref. [\onlinecite{spalek}], all
transitions are due to the interplay between band formation and
electron Coulomb correlation.

In 1978 Castellani, Natoli and Ranninger\cite{cnr} gave  a
realistic description of the complex magnetic properties and phase
diagram of V$_2$O$_3$, focusing on the peculiar magnetic structure
observed in the AFI phase, which breaks the high-temperature
trigonal symmetry and cannot be explained in terms of a
single-band Hubbard model. They realized that the introduction of
the extra degree of freedom represented by the degeneracy of t$_{2g}$-vanadium
orbitals was necessary to describe correctly the spin structure,
whose driving mechanism was the ordered pattern of
t$_{2g}$-orbitals throughout the whole crystal. This model was found to be in need of correction only more than 20 years later, when two experiments of non-resonant magnetic x-ray scattering\cite{paolasini} and
linear dichroism\cite{park} independently
demonstrated that the average spin value on each vanadium ion is
$S_V=1$, contrary to the previously supposed $S_V=1/2$.\cite{cnr}
These latter measurements stimulated a renewed interest in the
system, that led to a new picture for
its ground-state, where the two nearest V-ions are linked together
in a stable molecule with spin
$S_M=2$.\cite{mila12,dimatteo,tanaka} When considered together with the available sets
of x-ray diffraction data,\cite{dernier} and  polarized neutron
experiments,\cite{moon,weibao} this model predicts that the AFI phase is not magnetoelectric, as further confirmed
by two experiments.\cite{astrov,jansen} 

Yet, this seems
not to be in keeping with the detection of non-reciprocal effects in linear dichroism, in Ref. \onlinecite{goulon}, whose existence requires a magnetoelectric (ME) phase. Indeed, on the basis of diffraction data,\cite{dernier,moon,weibao} the magnetic space group (MSG) for the AFI phase of V$_2$O$_3$ turns out to be $P2/a+{\hat {T}}\{ {\hat {E}}|t_0 \} P2/a$, which is not magnetoelectric, because it explicitly contains both time-reversal (${\hat {T}}$) and inversion (${\hat {I}}$) symmetries. However, a necessary condition to reveal a non-reciprocal signal is that none of them be a symmetry operator.\cite{goulon} Thus, as better detailed in Ref. [\onlinecite{lindic}], the interpretation of Ref. [\onlinecite{goulon}] is in contradiction with the usually accepted MSG.

In this work we present a new experiment of vanadium K edge absorption and linear dichroism, with the aim of clarifying the physical picture and solve the controversy about the ground-state symmetry of V$_2$O$_3$. In order
to better compare with the results of Ref. [\onlinecite{goulon}],
we used the same single crystal, which is 2.8$\%$ Cr-doped. The
theoretical framework at the basis of this experiment is the one 
introduced in Ref. [\onlinecite{lindic}], where the possibility of
a structural origin for the observed dichroism was suggested and a
numerical prediction of the effect was carried out. As shown in the
next two sections, the present measurement confirms
that prediction and calls for a final experiment to check the effect of a ME annealing procedure\cite{noteme} when cooling the system down to its AFI phase.

\section{Experimental set-up.}

In the monoclinic low-temperature phase pure V$_2$O$_3$, as well
as the 2.8$\%$ chromium-doped sample used in the experiment,
crystallizes in the body-centered classical space group $I2/a$,\cite{dernier}
with two groups of four V ions in the unit cell, which are translationally equivalent (together with their oxygen environment) apart from the opposite direction of magnetic moments.\cite{moon,weibao} 
X-ray absorption spectra at V K edge (E$_o$=5465 eV) were
collected at the ID26 beamline of the European Synchrotron
Radiation Facility in fluorescence mode.
 The x-ray beam energies were selected by a double crystal
Si(220), fixed exit monochromator with an energy resolution of
about 0.3 eV at V K edge. The ID26 beamline worked in the so-called
gap-scan mode, that is tuning the maximum of the undulator emission to the actual
monochromator rocking curve.  Two mirrors provided an harmonic
free, small (diameter $\sim 300 \mu m$), intense x-ray spot size on
the sample, linearly polarized in the horizontal plane. The fluorescence intensity
emitted from the sample was collected in the total-fluorescence mode
by measuring the current from two photodiodes, mounted parallel to
the polarization of the incoming beam in order to minimize the
elastic contribution to the spectrum.

\begin{figure}
\epsfysize=75mm
\centerline{\epsffile{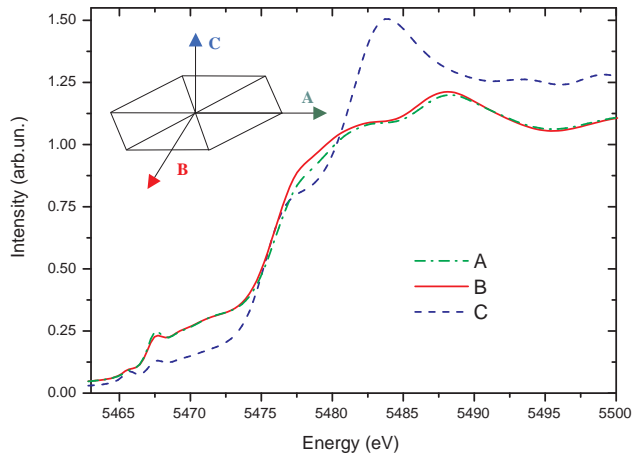}}
\caption{Experimental "in-plane" (A), (B) and "out-of-plane" (C) XANES spectra. A and B data are an average over 61 spectra; C is an average over 16 spectra.}
\label{fig1}
\end{figure}

The 2.8$\%$ chromium-doped V$_2$O$_3$ single crystal (about
$5\times 5 \times 2$ mm$^3$) was mounted in a closed cycle He
refrigerator, properly oriented with respect to the horizontal
linear polarization of the x-ray beam and cooled down to 80 K.
Absorption spectra were collected in quick-EXAFS mode (20 seconds
per scan) in the  energy range 5300-6300 eV (about 4000 points per
scan). Our geometrical setup allowed to have very narrow Bragg peaks. In order to remove them, several ($\simeq$60) spectra
were collected at different angles by rocking the sample ($\pm$ 2
deg.) around the vertical axis. With this procedure Bragg peaks were shifted and could be easily distinguished from the true spectral features. Averaging up all the scans then provided exceptional quality, low noise, XAS spectra suitable for accurate
quantitative studies. 
All data were collected for three orthogonal directions of polarization, ie, along the corundum c-axis, and in the hexagonal plane, along the monoclinic b$_m$-axis and orthogonal to it, as shown in Fig. \ref{fig1}. Here we analyze the data collected in the near-edge region.

\section{Results.}

At V K edge, the x-ray absorption
cross-section for the $i^{th}$ vanadium ion in the unit cell is
given by:

\begin{eqnarray}
\sigma_i=4\pi\alpha^2\hbar\omega\sum_n
|\langle\Psi_n^{(i)}|{\hat{O}}|\Psi_0^{(i)}\rangle |^2
\delta[\hbar\omega-(E_n-E_0)] \label{crsec}
\end{eqnarray}

The operator ${\hat {O}}\equiv {\hat {\epsilon}} \cdot \vec{r} (1
+ \frac{i}{2} \vec{k} \cdot \vec{r})$ is the usual
matter-radiation interaction operator expanded up to the
quadrupolar term, with photon  polarisation ${\hat {\epsilon}}$
and wave vector $\vec{k}$. $\Psi_0^{(i)}$ ($\Psi_n^{(i)}$) is the
ground (excited) state of the crystal and $E_0$ ($E_n$) its
energy. The sum is extended over all the excited states of the
system, $\hbar\omega$ is the energy of the incoming photon and
$\alpha$ is the fine-structure constant. The total cross-section
is obtained by summing up over all 8 ions in the unit cell: $\sigma=\sum_{i=1}^8 \sigma_i$. The
corresponding sum-rule for both structural and non-reciprocal
linear dichroism was already derived in Ref.
[\onlinecite{lindic}] in the framework of multiple scattering
theory. Here we just remind
that on the basis of the usually accepted
$P2/a+{\hat {T}}\{ {\hat {E}}|t_0 \} P2/a$ MSG,\cite{lindic}
linear dichroism in the hexagonal plane is expected to be at a maximum when one of the
two orthogonal polarizations is parallel to the monoclinic
b$_m$-axis, thus orthogonal to the glide plane, and is zero when
both polarizations are at $45^o$ degrees with the latter.

\begin{figure}
\epsfysize=75mm
\centerline{\epsffile{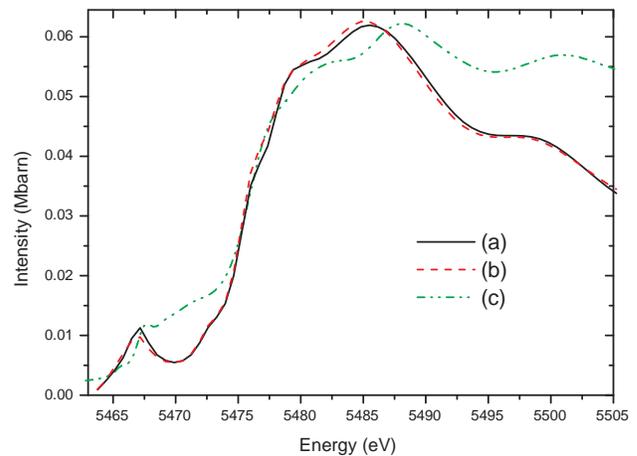}}
\caption{Theoretical and experimental "in-plane" XANES. (a) and (b) are the calculated A and B spectra in Mbarn. (c) is the experimental figure A normalized to the theory edge peak. }
\label{fig2}
\end{figure}

\begin{figure}
\epsfysize=75mm
\centerline{\epsffile{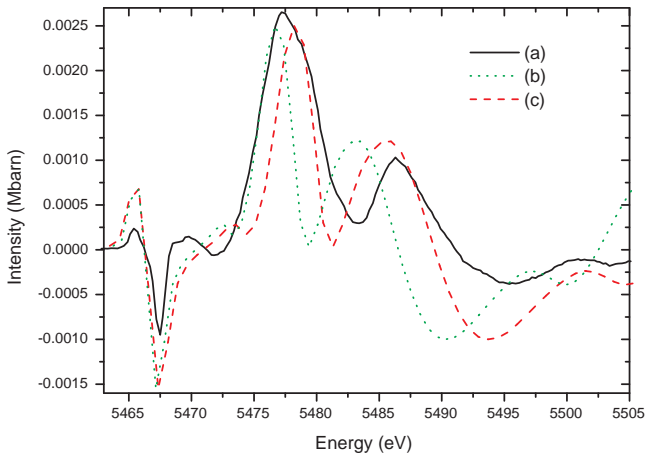}}
\caption{Experimental and theoretical "in-plane" linear dichroism. (a) is the experimental plot, (b) the theory, and (c) is the theory with an expanded energy scale (see text).}
\label{fig3}
\end{figure}

Figure \ref{fig1} shows the experimental data in the XANES region for the three polarization directions:
when the x-ray polarization is along the c-axis (C), dichroism with respect to the other two directions is apparent; less manifest, but still clear is the dichroism within the hexagonal plane, with the x-ray polarizations respectively parallel to the monoclinic
b$_m$-axis (B) and perpendicular to it (A). Such dichroism is explictly shown in Fig. \ref{fig3}(a).

In Fig. \ref{fig2}, instead, we have plotted a numerical 
simulation of the "in-plane" XANES, as calculated in the
framework of the multiple scattering (MS) theory within the muffin-tin
approximation, and superimposed the experimental curve (B) for a better comparison. 
As in Ref. [\onlinecite{lindic}], we chose a cluster
containing 135 atoms, {\it i.e.}, 54 vanadium and 81 oxygen atoms in
the correct ratio 2 to 3, having a radius of 6.9 \AA, enough to
get convergence with the cluster size. The spectrum calculated with the real
part of the Hedin-Lundqvist (HL) potential is convoluted with a
lorentzian function having an energy dependent damping $\Gamma(E)$,
derived from the universal mean free path curve by the relation
$\lambda(E) = (1/k_e)~E/\Gamma(E)$, where $E$ is the photoelectron
kinetic energy (in Rydbergs) and $k_e = \sqrt{E}$ its wave vector.
The agreement between experimental
measurements and numerical simulations is quite good, even though the energy distance between the peak at 5467 eV and the highest peak is underestimated in the theory (18 eV vs. 20.5 eV). This is probably a consequence of the potential adopted to describe the electron-atom scattering in the MS equations, whose HL exchange-correlation part\cite{ref} is known to be too attractive near the edge.\cite{benf}

This compression of the energy scale in the theoretical signal is highlighted in Fig. \ref{fig3}, where experimental and theoretical "in-plane" linear dichroisms are compared. In spite of it, all other main features of the experiment are well described by our calculation, as illustrated in Fig. \ref{fig3}(c), where the theoretical energy scale is expanded by a factor of 1.14 in order to better guide the eye: in this case all main peaks and valleys are correctly reproduced at almost all energies. The same happens for XANES in Fig. \ref{fig2}, apart from the experimental signal around 5470 eV, to be probably related to the upper Hubbard band,\cite{ezhov} which is missed by our single-particle approach. Even more important, no extra scale factors are needed to describe the relative intensity of linear dichroism and XANES spectra, as both experiment and theory show a ratio of about 3$\%$.
Finally, the calculated intensity is found to be of purely dipolar (E1-E1) origin, due to the structural monoclinic distortion, and not of magnetic origin.

All this is to be compared with the absence of structural linear dichroism
measured in Ref. [\onlinecite{goulon}], as derived from their Fig.
2, where the average signal in the two magnetic field
configurations is shown. The
apparent contradiction can be rationalized only in two ways, as suggested in Ref. [\onlinecite{lindic}]: either by assuming a transition to another monoclinic domain, subsequent to the ME annealing procedure, or by an accidental geometrical setup such that the two orthogonal polarizations were at 45$^o$ from the b$_m$-monoclinic axis. In this latter case there is no structural dichroism, as shown in Ref. [\onlinecite{lindic}].

In the present experiment, when the system was heated up to its PI phase and then cooled back to the AFI phase, without any ME annealing procedure, we found a transition to the same monoclinic domain.
Even though a proper explanation of this phenomenon is still lacking, it is interesting to note that all $2.8\%$ Cr-doped V$_2$O$_3$ single crystals used in resonant x-ray scattering experiments\cite{paolasini,paolasini2,alebomba} had the tendency to form one big monodomain, instead of a statistical mixture of the three possible.

Under the assumption of polarization at 45$^o$ from the b$_m$-axis, the appearence of a non-reciprocal signal in the AFI phase might find its origin in the ME annealing procedure, that might have driven the system to a nearby excited ME state that is known to exist from Ref. [\onlinecite{dimatteo}]. 
However, we would like to stress again that the relative intensity of the linear dichroism at the main peak is about 3$\%$ of XANES in all three cases, i.e., the present experiment, our theoretical calculation, and Ref. [\onlinecite{goulon}], thus strongly pointing towards a common origin of the three, given also the similarity in shape. In this respect, the possibility that the signal of Ref. [\onlinecite{goulon}] be of magnetic E1-E2 origin is in contradiction with the resonant x-ray scattering results of Paolasini {it et al.}\cite{paolasini2}: there it is shown that the $(10\overline{1})$ reflection, of E1-E1 nature, is 100 times more intense than the $(3\overline{1}1)$, of E1-E2 magnetic origin, which is moreover restricted to the pre-edge region ($\simeq 5$ eV) and not extended for more than 40 eV, like the linear dichroism of Fig. \ref{fig3}.
The factor 10 in the amplitude ratio of E1-E1 and E1-E2 signals for a single $V$-ion site is also confirmed by our ab-initio calculations, as well as by the practical absence of states with $d$ symmetry in the energy range 5474-5505 eV. 

In conclusion, the E1-E1 origin of this linear dichroism experiment is well supported
by a robust theoretical framework and calls for a further,
definitive, investigation of the ground state properties of
V$_2$O$_3$ in its low-temperature phase subsequent to a
magnetoelectric annealing.

----



\begin{thebibliography}{99}
\bibitem{rice}
D.B. McWhan, T.M. Rice, and J.P. Remeika, Phys. Rev. Lett. {\bf 23}, 1384 (1969)
\bibitem{mcwhan1}
D.B. Mc Whan and J.P. Remeika, Phys. Rev. B {\bf 2} 3734 (1970); D.B. McWhan, J.P. Remeika, S.D. Bader, B.B. Triplett, and N.E. Philips, Phys. Rev. B {\bf 7}, 3079 (1973); H. Kuwamoto, J.M Honig, and J. Appel, Phys. Rev. B {\bf 22}, 2626 (1980) and references therein.
\bibitem{dernier}
P.D. Dernier and M. Marezio, Phys. Rev. {\bf 2}, 3771 (1970)
\bibitem{moon}
R.B. Moon, Phys. Rev. Lett. {\bf 25}, 527 (1970)
\bibitem{spalek}
J. Spa{\l}ek, J. Sol. State Chem. {\bf 88}, 70 (1990)
\bibitem{cnr}
C. Castellani, C.R. Natoli and J. Ranninger, Phys. Rev. B {\bf 18}, 4945 (1978); ibid., 4967 (1978); ibid., 5001 (1978)
\bibitem{paolasini}
L. Paolasini, C. Vettier, F. de Bergevin, F. Yakhou, D. Mannix, A. Stunault, W. Neubeck, M. Altarelli, M. Fabrizio, P.A. Metcalf and J.M. Honig, Phys. Rev. Lett. {\bf 82}, 4719 (1999)
\bibitem{park}
J.-H. Park, L.H. Tjeng, A. Tanaka, J.W. Allen, C.T. Chen, P. Metcalf, 
J.M. Honig, F.M.F. de Groot, G.A. Sawatzky, {\it Phys. Rev.} B {\bf 61}, 11506 (2000)
\bibitem{mila12}
F. Mila, R. Shiina, F.-C. Zhang, A. Joshi, M. Ma, V. Anisimov, T.M. Rice, 
Phys. Rev. Lett. {\bf 85}, 1714 (2000)
\bibitem{dimatteo}
S. Di Matteo, N.B. Perkins, C.R. Natoli,  Phys. Rev. B {\bf 65}, 054413 (2002)
\bibitem{tanaka}
A. Tanaka, Jour. Phys. Soc. Jpn. {\bf 71}, 1091 (2002)
\bibitem{astrov}
D.N. Astrov, Sov. Phys. JETP, {\bf 11}, 708 (1960)
\bibitem{jansen}
S. Di Matteo, A.G.M. Jansen, Phys. Rev. B {\bf 66}, 100402(R) (2002)
\bibitem{goulon}
J. Goulon, A. Rogalev, C. Goulon-Ginet, G. Benayoun, L. Paolasini, 
C. Brouder, C. Malgrange, P.A. Metcalf, Phys. Rev. Lett. {\bf 85}, 
4385 (2000)
\bibitem{lindic}
S. Di Matteo, Y. Joly, C.R. Natoli, Phys. Rev. B {\bf 67}, 195105 (2003)
\bibitem{weibao}
Wei Bao, C. Broholm, G. Aeppli, S.A. Carter, P. Dai, T.F. Rosenbaum, J.M. Honig, P. Metcalf, S.F. Trevino,  Phys. Rev. B {\bf 58}, 12727 (1998)
\bibitem{noteme}
The ME annealing procedure consists in the application of external electric and magnetic fields when
cooling the sample through a phase transition. 
\bibitem{ref}
L. Hedin, B.I. Lundqvist, J. Phys. C, {\bf 4}, 2064 (1971)
\bibitem{benf}
C.R. Natoli, M. Benfatto, S. Della Longa, K. Hatada, J. Synchrotr. Rad. {\bf 10}, 26 (2003) and references therein
\bibitem{paolasini2}
L. Paolasini, S. Di Matteo, C. Vettier, F. de Bergevin, A. Sollier, W. Neubeck, F. Yakhou, P.A. Metcalf, and J.M. Honig, J. Electr. Spectr. \& Rel. Phen. {\bf 120}, 1 (2001)
\bibitem{ezhov}
S.Y. Ezhov, V.I. Anisimov, D.I. Khonskii and G.A. Sawatzky, Phys. Rev. Lett. {\bf 83}, 4136 (1999)
\bibitem{alebomba}
A. Bombardi, F. de Bergevin, S. Di Matteo, L. Paolasini, P.A. Metcalf, J.M. honig, Physica B {\bf 345}, 40 (2004)



\end{thebibliography}
\end{document}